# THE OPTICAL IMAGER "GALILEO" (OIG) *

**F. Bortoletto[1,2], S. Benetti[1], G. Bonanno[3], C. Bonoli[2], R. Cosentino[3], D'Alessandro[2], D. Fantinel[2], A. Ghedina[1], E Giro[2], A. Magazzu[1], C. Pernechele[2], C. Vuerli[4]**

[1]*Centro Galileo Galilei - TNG, La Palma, Canarie (E)*
[2] *Osservatorio Astronomico di Padova*
[3] *Osservatorio Astrofisico di Catania*
[4] *Osservatorio Astronomico di Trieste*

## 1. Introduction

The present paper describes the construction, the installation and the operation of the Optical Imager Galileo (OIG), a scientific instrument dedicated to the *'imaging'* in the visible.

OIG was the first instrument installed on the focal plane of the Telescopio Nazionale Galileo (TNG) and it has been extensively used for the functional verification of several parts of the telescope (as an example the optical quality, the rejection of spurious light, the active optics and the tracking), in the same way also several parts of the TNG informatics system (instrument commanding, telemetry and data archiving) have been verified making extensive use of OIG.

This paper provides also a frame of work for a further development of the imaging dedicated instrumentation inside TNG.

OIG, coupled with the first near-IR camera (ARNICA), has been the 'workhorse instrument' during the first period of telescope experimental and scientific scheduling.

## 2. The Instrument and the Schedule

OIG is the TNG CCD camera for direct imaging for the optical region between 0.32 and 1.1 microns proposed in 1992 to the former Italian Astronomical Research Council (CRA) and documented in Ref. 1-2. It is mounted on the Nasmyth-A adapter interface and is normally illuminated by the f/11 light beam directly coming from the TNG tertiary mirror into the focal plane (plate-scale: 5.36 *arcsec/mm*) with no other optical element in front of the CCD except those on the filter wheels and the cryostat window. Alternatively, OIG can be fed by a f/32 beam coming from the Adaptive Optics Module (ADOPT).

The optical window on the cryostat acts also as a field flattener in order to compensate the focus mismatching at large scale produced by the telescope intrinsic field-curvature. In this configuration OIG can exploit, without any degradation, the best observing conditions available in La Palma.

OIG was designed to host a variety of CCD chips or mosaics covering a field of view up to 10 *arcmin*. At the moment it is equipped with a mosaic of two thinned and back-illuminated EEV 42-80 CCDs with 2048 x 4096 pixels each (pixel size of 13.5 microns). In this configuration (see Fig. 2.1) the resulting pixel scale is 0.072 *arcsec/pix* for a total field of view of about 4.9 x 4.9 *arcmin*. The separation between the two chips is of about 2.8 *arcsec* and the misalignment between the two detectors is of about 0.07 degrees.

An alternative version of OIG, based on a single SITEK 2Kx2K CCD with a pixel size of 24 microns, is in preparation in order to better cover the I band where the EEV42-80 CCD is strongly affected by interference fringing (see Fig. 3.2). Figures 2.1 and 2.2 show OIG in several phases during the implementation at the telescope.

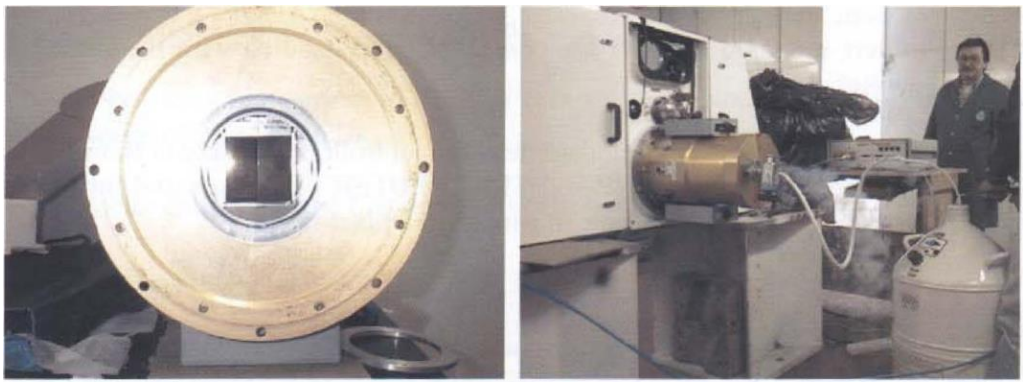

Fig. 2.1 — The OIG cryostat with the detector mosaic currently in use (left side).
The first mounting of OIG at the Nasmyth-A instrument adapter (October 1998. right side).

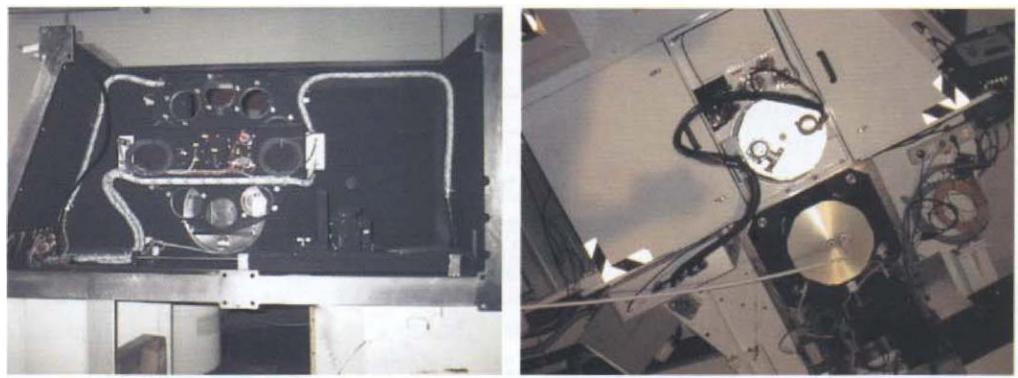

Fig. 2.2 — The OIG filter wheel inside the Nasmyth-A instrument adapter (left side).
OIG and ARNICA in operation at the telescope (right side).

The main characteristics of the instrument and a comparison with the imaging capabilities of the Low Resolution Spectrograph (DOLORES) and with the near infrared camera (NICS) are available in Tab. 2.1.

**Tab. 2.1 - Main characteristics of 01G, DOLORES and NICS in imaging mode**

|  | OIG | DOLORES | NICS |
|---|---|---|---|
| Detector(s) | 2x EEV CCD42-80 | LORAL 3edges | ROCKWELL |
| Format | 2x 2048x4096 | 2048x2048 | 1024x1024 |
| Pixel-size (μm) | 13.5x13.5 | 15x15 | 18.5x18.5 |
| Sampling *(aresec/pix)* | 0.07 | 0.27 | 0.25/0.13 |
| Field of View *(arcmin)* | 5x5 | 9.4x9.4 | 4.3x4.3/2.2x2.2 |

The readout and control system of OIG has been specifically constructed having in mind the similar needs for two other instruments, namely the High and the Low Resolution Spectrographs (SARG and DOLORES respectively), and the needs for the service cameras of TNG (trackers and Shack-Hartmann analyzers) and for the high speed wave front analyzer of the Adaptive Optics System (ADOPT). Presently all the CCD controllers mounted in TNG are compatible and handled with the same basic software (see Ref. 3).

A schematic view of the TNG's standard detector controllers structure is shown in Fig. 2.3; the controller is based on DSP and TRANSPUTER processors and, in its basic version, can process four CCD outputs in parallel at a pixel-rate of 0.1 MHz and with a readout noise basically limited by the detector itself. Each controller, or group of controllers, is handled through a dedicated VME crate providing dual-port buffering of each picture. A new version of this controller, with a faster pixel processor, is under implementation.

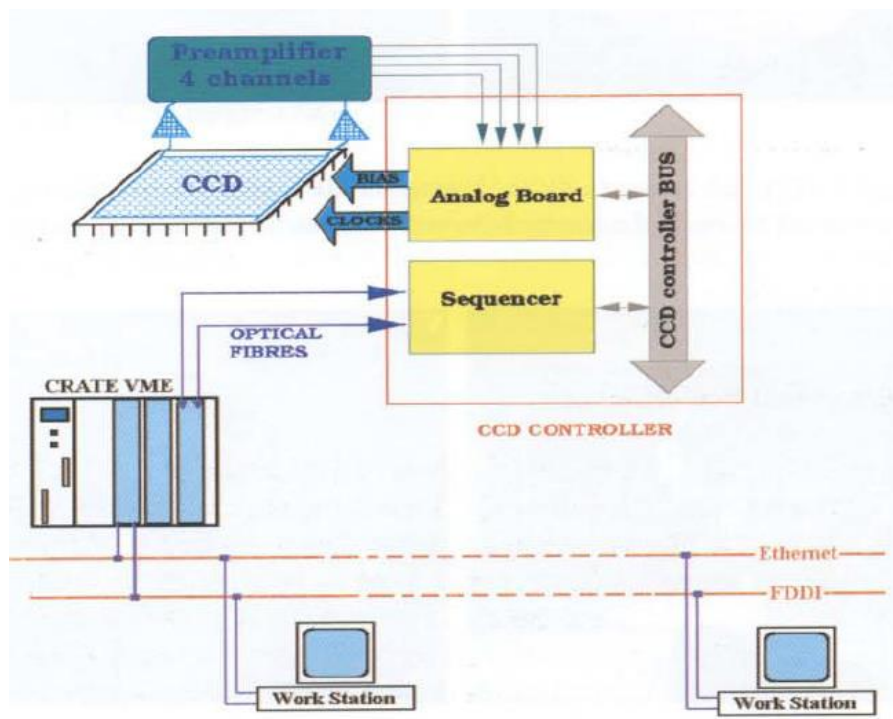

Fig.2.3 -The structure of a typical CCD controller inserted on the
TNG control net.

The general handling software for this instrument, in particular the IDL based user-interface, has been developed during the TNG commissioning and has been used as a paradigm for the development of several other interfaces (pointing and guiding, active-optics, several instruments).

The implementation schedule of this instrument has been:

- 1993, start of a contract for the provision of a CCD wafer-batch to the LORAL company (USA);

- 1993, start of a contract for the construction of the standard TNG detector controller to the LES company (ITALY);

- 1995, start of a contract to the Steward Observatory for the thinning and back-side processing of the LORAL chips;

- 1995, start of a contract to the EEV company (England) for the provision of a series of 2Kx4K CCDs;

- 1996, delivery of the first detector controller;

- 1997, delivery of two good quality CCDs by Steward Observatory;

- 1998, first light of OIG in November;

- 1999, delivery of the last CCD by EEV.

The costs involved in the implementation of the system are inside the budget foreseen at the start of the activities.

## 3. The Instrument Performance

The Relative Quantum Efficiency (RQE) of the detectors mounted inside OIG is shown in Fig. 3.1; in the same figure the RQE curves for the two thinned LORAL chips available for the low resolution spectrograph are shown. In the LORAL case the back-side processing is different for the two chips in that one is optimized for the red side of the spectra while the second is for the blue side and that is well evidenced by Fig. 3.1. Although the EEV CCD is processed with an antireflective coating optimized for the blue side of the spectra, nevertheless the LORAL chips seem to provide some more efficiency on the extreme blue at the expense of a lower peak RQE.

The benefit of a good efficiency, especially on the critical blue region below 0.4 microns can be seen on the photometric color terms obtained observing standard stars with OIG (see Tab. 3.1).

As an example of the actual camera performance the tested limit magnitude of OIG with an average seeing condition (0.7 *arcsec*) is of 25.5 in the V band at a signal to noise ratio of 5 and with an exposure time of 30 minutes.

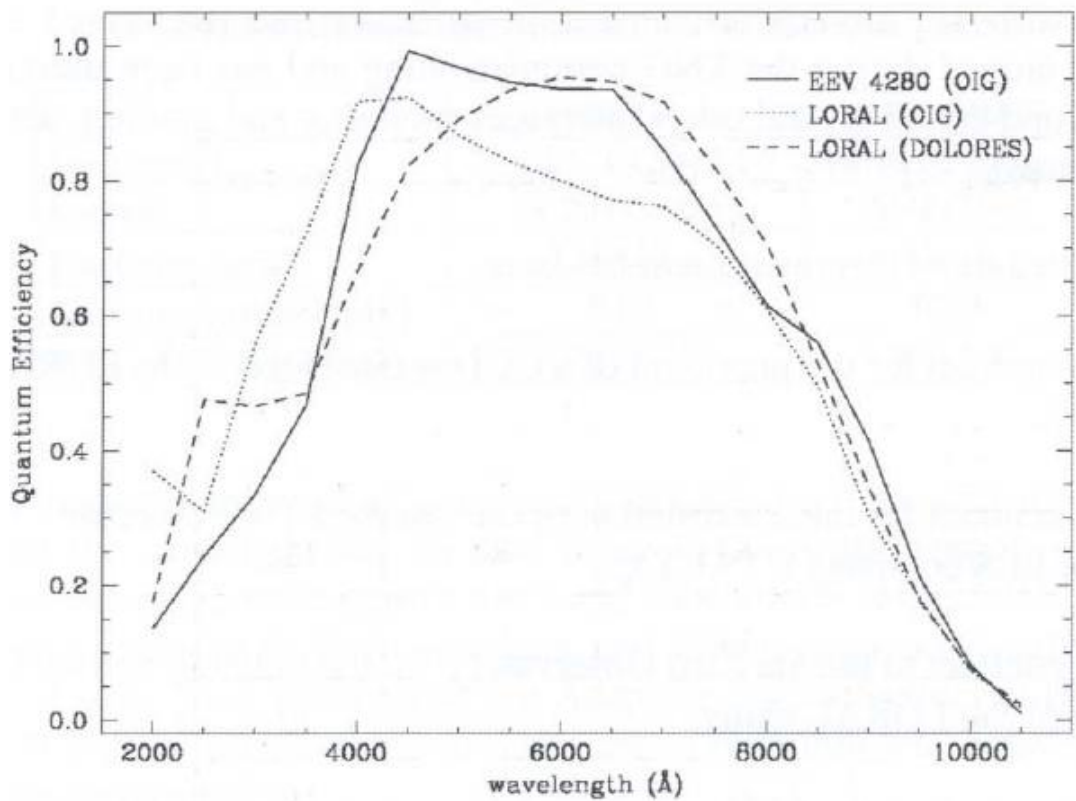

Fig. 3.1 - The quantum efficiency diagrams for three kinds of CCDs in use at the telescope.

The photometric filter system mounted inside OIG is presently based on two filter sets: the standard Johnson-Cousins set and the Gunn set.

The main parameters of the filters can be seen in Tab. 3.2; more information is available in http://www.tng.iac.es/instruments/oig/oig.html .

Tab. 3.1 — The 01G colour-indexes calibration for the two EEV detectors mounted inside OIG

| COLOUR INDEX | CCD1 | CCD2 |
|:---:|:---:|:---:|
| ΔU | 24.556 + 0.11x(U-B) | 24.654 + 0.14x(U-B) |
| ΔB | 26.282 + 0.10x(B-V) | 26.266 + 0.09x(B-V) |
| ΔV | 26.156 — 0.06x(B-V) | 26.159 - 0.06x(B-V) |
| ΔR | 26.031 + 0.14x(V-R) | 26.033 + 0.13x(V-R) |
| ΔI | 25.555 + 0.02x(R-I) | 25.568+ 0.004x(R-I) |

The cosmetic quality of the EEV set of chips can be estimated from the series of night sky flats shown in Fig. 3.2 covering the U, V and I band. In the extreme blue side the disuniformity of both chips is limited by the residual of the back-side after-thinning processing while in the extreme red side it is limited by fringing, as usual for several thinned CCDs.

Tab. 3.2 — The main parameters of the Johnson and Gunn photometric systems mounted on OIG

| Filter | Peak WL. (nm) | Centr. WL. (nm) | Peak Tr. (%) | FWHM (nm) |
|--------|---------------|-----------------|--------------|-----------|
| U | 367 | 361 | 67 | 60 |
| B | 445 | 436 | 67 | 103 |
| V | 523 | 533 | 84 | 93 |
| R | 592 | 625 | 82 | 128 |
| g | 493 | 518 | 82 | 103 |
| r | 638 | 659 | 86 | 81 |
| i | 828 | 825 | 85 | 192 |
| z | 870 | ~900 | 90 | ~140 |

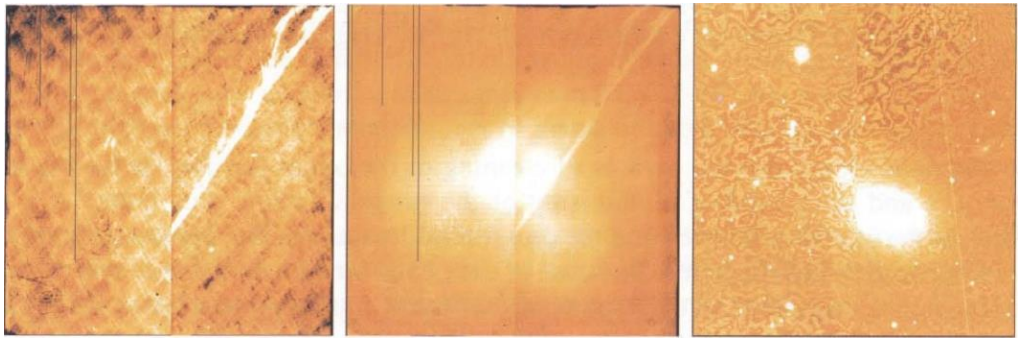

Fig. 3.2 A series of three OIG flats made on the sky with the Johnson U, V filters and the Gunn *i* band filter respectively (left to right).

The photometric history of OIG from the start of operations till now is well represented by the diagrams reported in Fig. 3.3. They show the magnitude calibration obtained from standard stars regularly observed during the last two years in the Johnson bands and in the *i* band.

The empty and the filled dots correspond to the chip l and 2, respectively, mounted in OIG. A comparison with similar measures for the Low Resolution Spectrograph and for the ESO's EFOSC-2 spectrograph (see the dashed and dot-dashed horizontal lines) is also reported. These measurements put in evidence the good performance of the instrument, especially in the blue region.

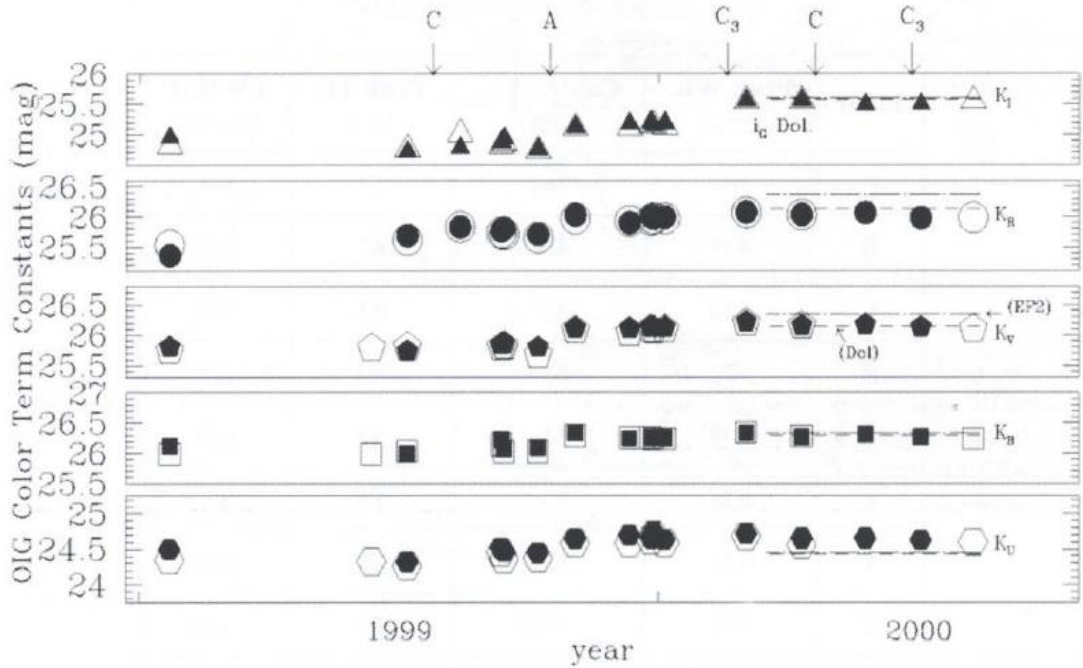

Fig. 3.3 — A comparison between the sensitivity of OIG (filled and empty dots), the low dispersion spectrograph DOLORES (horizontal dashed line) and ESO's EFOSC2 imager/spectrograph (horizontal dot-dashed line) cover the U, B, V, R and i bands from the bottom.

The diagrams report also the period of cleaning of the optics (arrows on the top with letter 'C') and the period of aluminising of the primary (arrow plus letter 'A'), the effects on the overall efficiency and the progressive loss due to the dust accumulation is well evidenced by the diagrams. The case of the *i* band curve is peculiar in that the gain of about half a magnitude in correspondence of the `C3' mark (cleaning of the tertiary mirror) is primarily due to the change of the *i* filter with a more efficient one made in the same period.

An example of the good performance of the system and its ability to exploit the moments of good seeing is given in the two exposures shown in Fig. 3.4. At the left side there is a composite picture of M57 nebulae obtained combining three 40 seconds exposures on the B, V and R filters with a resulting FWHM of 0.55 *arcsec,* the galaxies field at the right side has been taken in the Gunn *r* band with an exposure of 900 sec in april 2000. The average FWHM is of 0.46 *arcsec* and the limit magnitude is of about 25.5 .

## 4. What New Imaging at TNG?

An evident problem concerned with the direct imaging at the two naked F/11 Nasmyth foci is the difficulty in covering efficiently the allowable field. This is well evident from Tab. 2.1, where one can see that the PSF sampling, with a naked array and making use of the typical detector pixel size today commercially available, is redundant even during very good observing conditions.

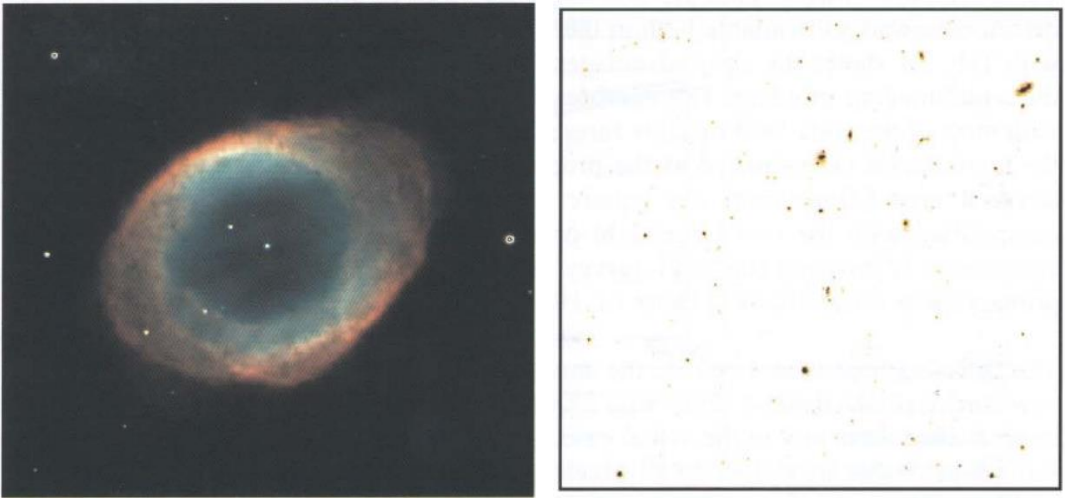

Fig. 3.4 — At the left side the M57 nebulae in Lyrae, at the right side a deep field of galaxies.

In the OIG case the detector pixel size is of 13.5 microns and the resulting scale is of 0.07"/pixel thus producing a considerable waste of detector pixels. Moreover the foreseen trend for the commercially available detectors (both in the visual and in the near IR) is to increase the number of pixels at the expense of the pixel size (recently EEV announced a new version of their large area EEV42-80 chip with a pixel size of about 7.5 microns, reaching a total number of pixels of about 4Kx8K making use of the same area of silicon).

As a simple example of the uncomfortable situation of the scientific imaging at the Nasmyth foci, the simple extension of OIG to cover the same field already covered by DOLORES would require 8 chips instead of the two presently installed.

TNG has been conceived from the start of the project (see Ref. 5) with the possibility of implementing a prime focus. Several mechanisms have been studied, built and verified during the telescope commissioning phase in order to facilitate the operations involved in mounting and dismounting a prime-focus instrument.

Tab. 4.1 — Main parameters for three different imaging configurations at the TNG prime-focus. The VISUAL case is based on a configuration with a final F-ratio of 2.2 while the NIR configuration is based on a 2.7 case. On the two mosaics some blind chip-to-chip gap must be considered. The *merit factor Q* is expressed as the product of the collecting area times the covered field (square-meters times square-degrees)

| MODE | DETECTOR SIZE (pixels) | PIXEL SIZE (µm) | MOSAIC FORMAT (chips) | TOTAL PIXELS | SAMPLING SCALE (arcsec/pix) | FIELD OF VIEW (arcmin) | MERIT FACTOR Q |
|---|---|---|---|---|---|---|---|
| VISUAL | 4Kx8K | 7.5 | 4x2 | 256 Mp | 0.2 | 55x55 | 10.3 |
| NIRI | 2Kx2K | 18 | 1x1 | 4 Mp | 0.39 | 14x14 | 0.66 |
| NIR2 | 2Kx2K | 18 | 2x2 | 16 Mp | 0.39 | 30x30 | 3 |

Table 4.1 shows what is possible to obtain at the TNG primary with the most advanced detectors nowadays available both in the visual and near-infrared bands, a comparison with Tab. 2.1 shows the clear advantages offered by the prime focus when addressing the wide imaging problem. The numbers are competitive, especially in the NIR side, with most of the wide field imagers foreseen in the northern hemisphere; as an example the merit factor Q, estimated as the product of the telescope area times the detector covered area (dimensions are square meters and square degrees), makes OIG competitive with the two large field projects in the visual band where the Italian community is involved (the VST survey telescope has a Q factor of 6.8 and the LBT primary focus imager has a Q factor of 10.24).

The detectors here considered are the announced EEV chip with 4Kx8K pixels and the new Rockwell HAWAII-2 array with 2Kx2K pixels. Among the various critical points to be studied there are, in the visual case, the decrease in dynamics due to the reduced saturation charge in a such small pixels, while in the NIR case, there is surely the problem of the reduction of thermal background (especially in the mosaic case, NIR2). The latter problem, in an instrument working without a cold pupil stop and with such a big angle of view, involves the study of an efficient baffling and shielding system.

The design of the prime-focus optical corrector (PFC) can be also moderately adapted (see Tab. 4.1 and Ref. 4) to provide the best match between field and PSF sampling.
An example of PFC design for the two NIR cases considered in Tab. 4.1 is given in Fig. 4.1; both designs are based on a set of four fused-silica optical elements with two aspheric surfaces each. Although several other combinations can be studied (see Ref. 4) the wide area corrector presented at the right side of Fig. 4.1 already shows a good optical performance (80% EE inside 18 microns, see also Fig. 4.2) and a good compactness (265 mm of entrance aperture and 420 mm of length).

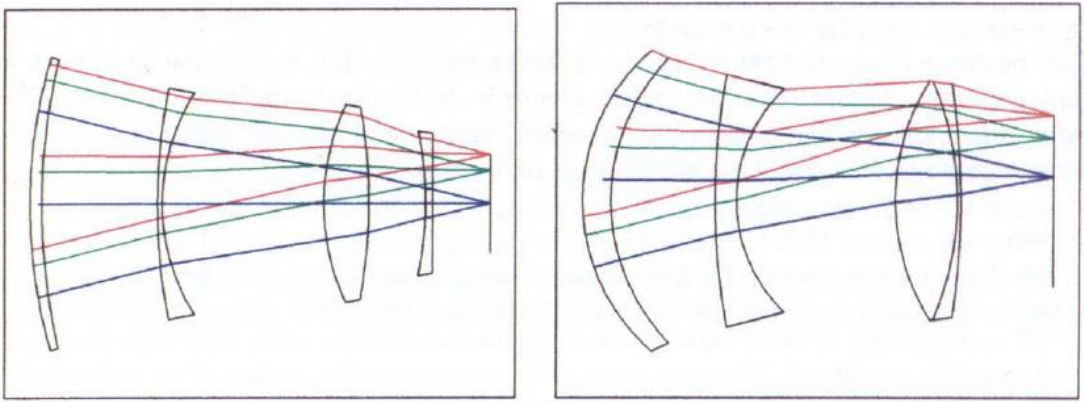

Fig. 4.1 - Two examples of prime-focus corrector, a single 2Kx2K NIR detector at left and a mosaic made of four 2Kx2K NIR detectors at right.

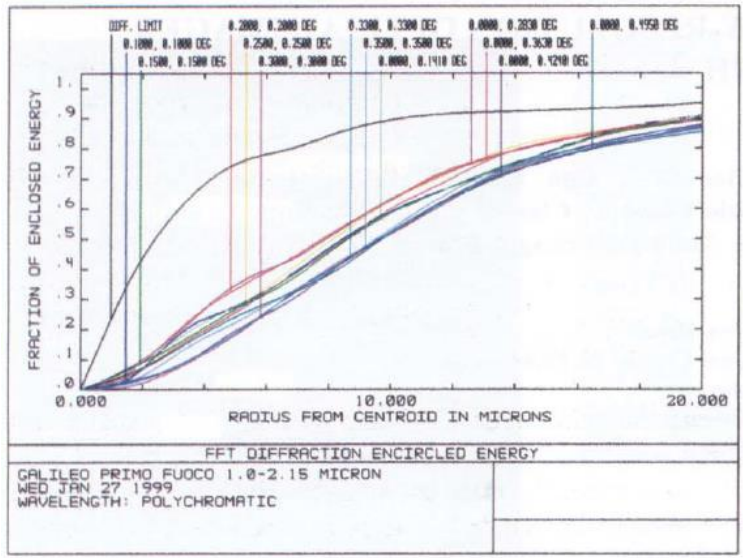

Fig. 4.2 — Encircled Energy (EE) curves for the mosaic corrector (right side of Fig. 4.1).

## 5. Conclusions

In this paper a description of the construction and of the actual performance of the TNG visual imager (OIG) has been given. The impact of this instrument in several areas (informatics, service-cameras, telescope commissioning) independently of the pure scientific operation has been evidenced.

A preliminary discussion on the future possibilities in order to increase the imaging capabilities of TNG has also been provided.